\documentclass[%
 reprint,
 amsmath,amssymb,
 aps,
]{revtex4-2}

\usepackage{graphicx}
\usepackage{dcolumn}
\usepackage{bm}
\usepackage{textcomp}
\usepackage{natbib} 
\usepackage{float}            
\usepackage{hyperref}

\usepackage{textcomp}
\usepackage[outdir=./]{epstopdf}
\usepackage{tabularx} 
\usepackage{times}
\usepackage{wrapfig}
\usepackage{xcolor}
\usepackage{caption}
\usepackage{subcaption}
\usepackage{adjustbox}
\usepackage[most]{tcolorbox}
\usepackage{listings}
\usepackage{silence}
\usepackage{wasysym}
\usepackage{titlesec}
\begin{document}

\preprint{APS/123-QED}

\title{Rotation of crystal seed during early stages of growth reveals the anisotropy of glass matrix}

\author{R. Thapa}
\email{rat422@lehigh.edu}
\affiliation{Department of Material Science and Engineering, \\
Lehigh University, Bethlehem, PA 18015, USA}%

\author{E. Mustermann}
\email{rat422@lehigh.edu}
\affiliation{Department of Material Science and Engineering, \\
Lehigh University, Bethlehem, PA 18015, USA}%

\author{H. Jain}%
\email{h.jain@lehigh.edu}
\affiliation{Department of Material Science and Engineering, \\
Lehigh University, Bethlehem, PA 18015, USA}%

\author{V. Dierolf}%
\email{vod2@lehigh.edu}
\affiliation{Department of Physics, \\
Lehigh University, Bethlehem, PA 18015, USA}%

\author{M. E. McKenzie}
\affiliation{Corning Incorporated, Science and Technology Division, \\
Corning, NY 14831, USA}
\date{\today}

\begin{abstract}
Rotation of crystal seed during the early stages of growth in a glass matrix has been observed due to some torque, contradicting the expectations from the  isotropic, uniform structure of the surrounding amorphous matrix. We establish an atomistic origin of this new phenomenon from molecular dynamics simulations using LiNbO\textsubscript{3} and LiNbO\textsubscript{3}-SiO\textsubscript{2} glasses as model systems. Effectively, it arises due to non-uniform forces on the seed from the surrounding glass, which appears inhomogeneous and anisotropic on the scale of glass-crystal interface. The seeded crystal growth (SCG) at higher temperatures amplifies this effect due to enhanced atomic dynamics. Silica, when added to LiNbO\textsubscript{3} glass, reduces the crystal growth rate due to increased viscosity and restricted atomic mobility across the growth interface, but has minimal effect on the crystal rotation. These findings challenge a general assumption that glass is an isotropic material, especially during the early stage of its crystallization, and provide insights for tailoring the microstructure of widely used glass-ceramics.

\end{abstract}

\maketitle

\section{\label{sec:level1}Introduction}
Nucleation and growth are two fundamental, omnipresent phenomena in any crystal growth technique used to crystallize glass into glass ceramics. The former is often described using classical nucleation theory (CNT), which, although imperfect, offers a straightforward physical understanding of the main mechanisms involved~\cite{CORMIER2014}. To address the shortcomings of CNT, several refined thermodynamic approaches have been proposed, such as the generalized Gibbs model~\cite{SCHMELZER2004, Schmelzer2006}, which incorporate more detailed considerations of interfacial energy and critical cluster size. However, these thermodynamic frameworks, while powerful in predicting macroscopic behavior, do not capture the microscopic atomistic pathways or local structural fluctuations that drive nucleation and growth. To explore these microscopic mechanisms, molecular dynamics (MD) simulations are employed. Unfortunately, the timescales required for nucleation and subsequent crystal growth often exceed what is accessible even with today’s most powerful supercomputers. To overcome this bottleneck, a variety of enhanced sampling and specialized simulation techniques have been developed. The most common methods include enhanced sampling approaches~\cite{TORRIE1977, Laio2002,Vlasson2016,Niu2018}, the free energy seeding method~\cite{Lodesani2020}, the persistent embryo method~\cite{Yang2018}, and the crystal seeding method~\cite{Sun2022, SUNJNCSX, RAJ2023}. For our system, where the critical nucleus size is already known from previous work~\cite{RAJ2023}, we chose to apply the seeded crystal growth technique to efficiently study the crystallization process.

The creation of single crystal inside glasses offers a wide range of applications in photonics, optoelectronics, and quantum memory devices. Although laser induced crystallization aims to achieve this by locally heating the glass to form a seed that is then extended along the laser scanning direction, it does not always result in single crystals. Veenhuizen {\it et. al}~\cite{VEENHUIZEN2017} demonstrated that the formation of single crystals depends sensitively on parameters such as scanning speed and power density, which effectively control the local thermal environment. By systematically varying the temperature or heating profiles, we can explore how different thermal conditions influence crystallization pathways - such as whether growth proceeds by layer-by-layer advancement, dendritic extension, or through the competition of multiple nuclei ultimately determining the structural quality of the resulting crystal. Very recently, it has been reported that the crystal seed rotates during early stages of laser crystallization in Sb\textsubscript{2}S\textsubscript{3} using CW laser, e-beam heating and X-ray~\cite{musterman2023structure}. The crystal rotation that ceases after a few seconds is independent of the heating source and has been seen in laser irradiation, e- beam heating, and X-ray exposure. The fundamental inquiry arising from this observation pertains to why crystal seed rotates despite being surrounded by an isotropic glass matrix. It could be argued that the glass surrounding the seed cannot be considered isotropic at that length scale. Although the effect of interface induced ordering of the liquid in the crystal growth rate has been reported before~\cite{Freitas2020}, there has been no study of the effect of the glass on crystal seed rotation. In this study, our objective is to investigate the factors that contribute to the rotation of the crystal seed during SCG. Since laser irradiation causes significant density fluctuations in and around the focal region, the seed nucleus may be surrounded by a loose network of atoms in or near the molten state. 

Atomic scale understanding of the seeded crystal growth (SCG) in LiNbO\textsubscript{3} (LNO) using molecular dynamics simulation has been previously reported~\cite{SUNJNCSX, RAJ2023}. In that study, we focused on understanding the growth of LNO crystal seed embedded in a glass of the same composition under isothermal-isobaric treatment below the theoretical melting point~\cite{RAJ2023}. In this work, we attempt to understand SCG in LNO and lithium niobosilicate (LNS) glasses and the origins of observed rotations of a seed. This understanding is extended to study the effect of the surrounding glass on the observed rotation of the seed.


\begin{table}[h!]
    \centering
    \begin{tabular}{|c|c|c|}
        \hline
        Name & $\rho\, (g/cc)$ & MQ time (ns)  \\ \hline
        I    & 4.08  & 1.9  \\ 
        II   & 4.08  & 2.1  \\ 
        III  & 4.08  & 2.3  \\ 
        IV   & 4.08  & 2.3  \\ 
        V    & 4.08  & 2.5  \\ 
        VI   & 4.08  & 2.7  \\ 
        VII  & 4.08  & 3.3  \\ 
        VIII & 4.30  & 1.7  \\ \hline
    \end{tabular}
    \caption{Nomenclature and details of melt-quench cycle for various LiNbO$_3$ models created for the study.}
    \label{tab:model_details}
\end{table}

\section{\label{sec:level2}Starting Conditions }
\subsection{\label{sec:level2A}Simulation details}
\begin{figure*}[!ht]
            \begin{minipage}[b]{.95\textwidth}
			\includegraphics[width=0.93\textwidth]{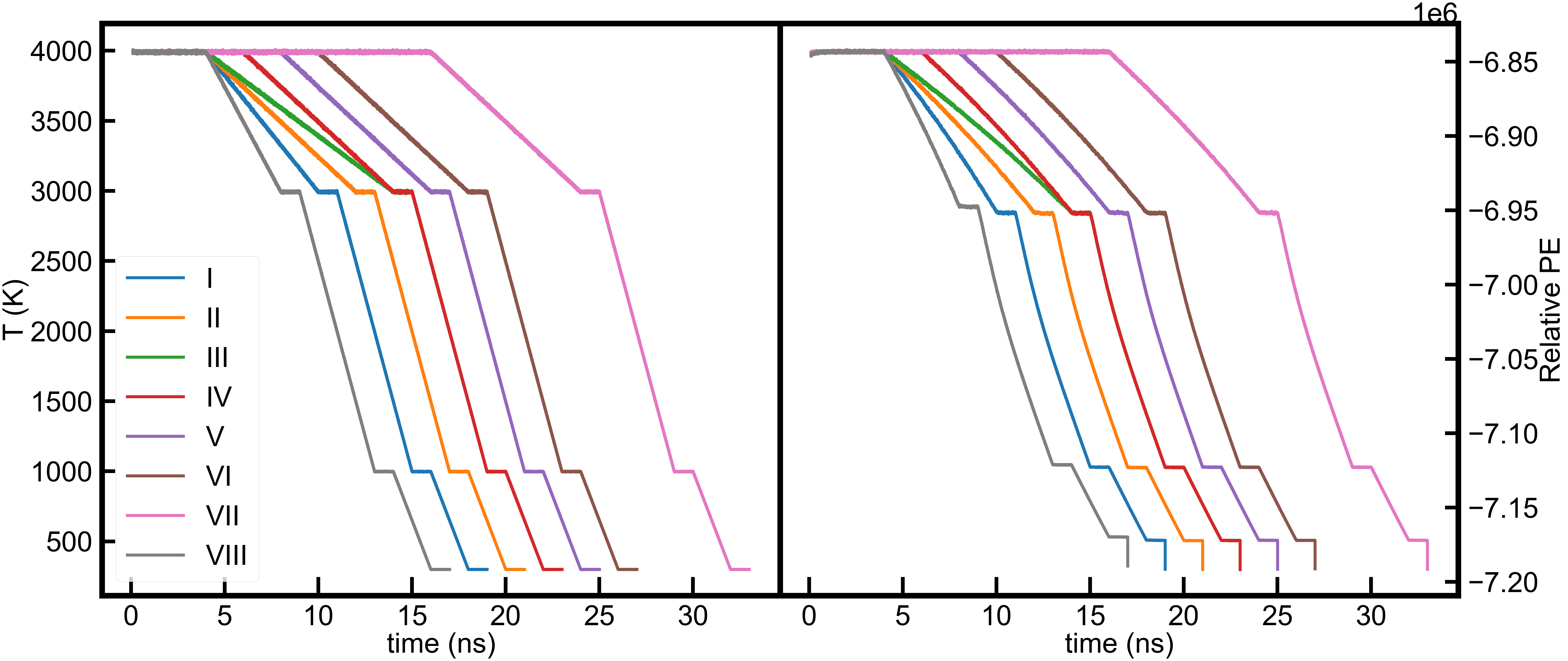}
			\caption{Time evolution of temperature and potential energy during the melt quench cycle for all models created.}
                \label{fig:t_PE_evol}
            \end{minipage}
\end{figure*}

We have created models of LNO glass with about half a million atoms. The system includes a single spherical seed, 15 $\mathring{A}$ in radius, at experimental density using the traditional melt-quench technique. The pairwise interaction for this simulation is a combination of a long-range Coulomb term and a short-range Buckingham term, with partial effective charge. The overall interaction takes the form:

\begin{equation}
    V_{ij} \left( r_{ij} \right) = \frac{Z_iZ_je^2}{4\pi\epsilon_0r_{ij}} + A_{ij}exp\left(-\frac{r_{ij}}{\rho_{ij}}\right) - \frac{C}{r_{ij}^6}
\end{equation}
, where $Z_i$ and $Z_j$ are the partial charges of atoms $i$ and $j$, and $r_{ij}$ is the interatomic distance between atom $i$ and $j$. The empirical parameters for the Buckingham potential were taken from Sun {\it et al.}~\cite{Sun2022}. The discrepancy of the Buckingham potential at short interatomic distances has been corrected using a correction factor~\cite{Du2019}. Since 15 $\mathring{A}$ seeds are larger than the critical nucleus size~\cite{RAJ2023}, we chose the seeds placed in the models to be 15 $\mathring{A}$ in size to model crystal growth and rotation during SCG, if any. The liquid melt was created, outside the seed, at 4000 K and was gradually cooled to room temperature with an equilibration cycle between the subsequent cooling cycle for a total simulation time ranging from 1.70 ns to 3.30 ns. The details of the simulation for all LiNbO$_3$ models are given in Table.~\ref{tab:model_details}. The time evolution of the temperature and potential energy during the melt-quench cycle is shown in Fig.~\ref{fig:t_PE_evol}. Selective dynamics was employed to keep the seeds fixed during the entire MQ cycle, while moving the atoms outside the seeds, in order to generate a seeded glass. All melt-quench and SCG molecular dynamics simulations were performed using the open source LAMMPS package~\cite{LAMMPS}. 

Two lithium niobosilicate (LNS) glasses, 98LiNbO\textsubscript{3}-2SiO\textsubscript{2} (LNS2) and 94LiNbO\textsubscript{3}-6SiO\textsubscript{2} (LNS6), with one LNO crystal seed and made up of about half a million atoms were also created to study the effect of silica on the crystal growth rate and the observed rotation during SCG. The total melt-quench time for both the glass were 2.3 ns. Once the melt-quench cycle was complete, the seeded glasses were annealed to simulate a seeded crystal growth. The seeded glass structure was passed on to an NPT ensemble (3000 K , 1 atm) allowing for atomic dynamics of all atoms present, seeded crystal and glass. NVT ensemble (3000 K) was also simulated to confirm whether the observed growth and rotation were artifacts of the simulation of the NPT ensemble. 

For both the LNO and LNS glasses, the time evolution of the crystal seed that grows and the residual glass has been tracked using a machine learning clustering algorithm based on static and dynamic structural factors~\cite{RAJ2023}. Visualizations have been created using the OVITO 3D visualization software~\cite{Stukowski_2010}.


\subsection{\label{sec:level2B}Crystal rotation quantification}
We developed a technique to quantify the crystal rotation observed during SCG. The algorithm for calculating the rotation angles during the SCG is as follows:
\begin{itemize}
    \item Consider the atoms in the crystal seed placed initially. Identify the position vector for each atom at time $t=0$, $\vec{r_i}(0)$, about the center of mass. 
    \item In each subsequent snapshot at time $t$, calculate the vector for each atom relative to the center of mass and call it $\vec{r_i}\, (t)$. 
    \item The rotation of an atom $i$ at time $t$ during the SCG, $\theta_i(t)$, is the angle  between the vectors $\vec{r_i}\, (0)$ and $\vec{r_i}\, (t)$. The rotation of the seed at time $t$ is given by: 
    
    \begin{equation}
        \theta_t = \frac{1}{N}\sum_{i=1}^N \theta_i
    \end{equation}   
    , where the sum runs over all atoms in the seed being considered. 
\end{itemize}


\begin{figure*}[!ht]
            \begin{minipage}[b]{.47\textwidth}
			\centering
			\includegraphics[width=0.93\textwidth]{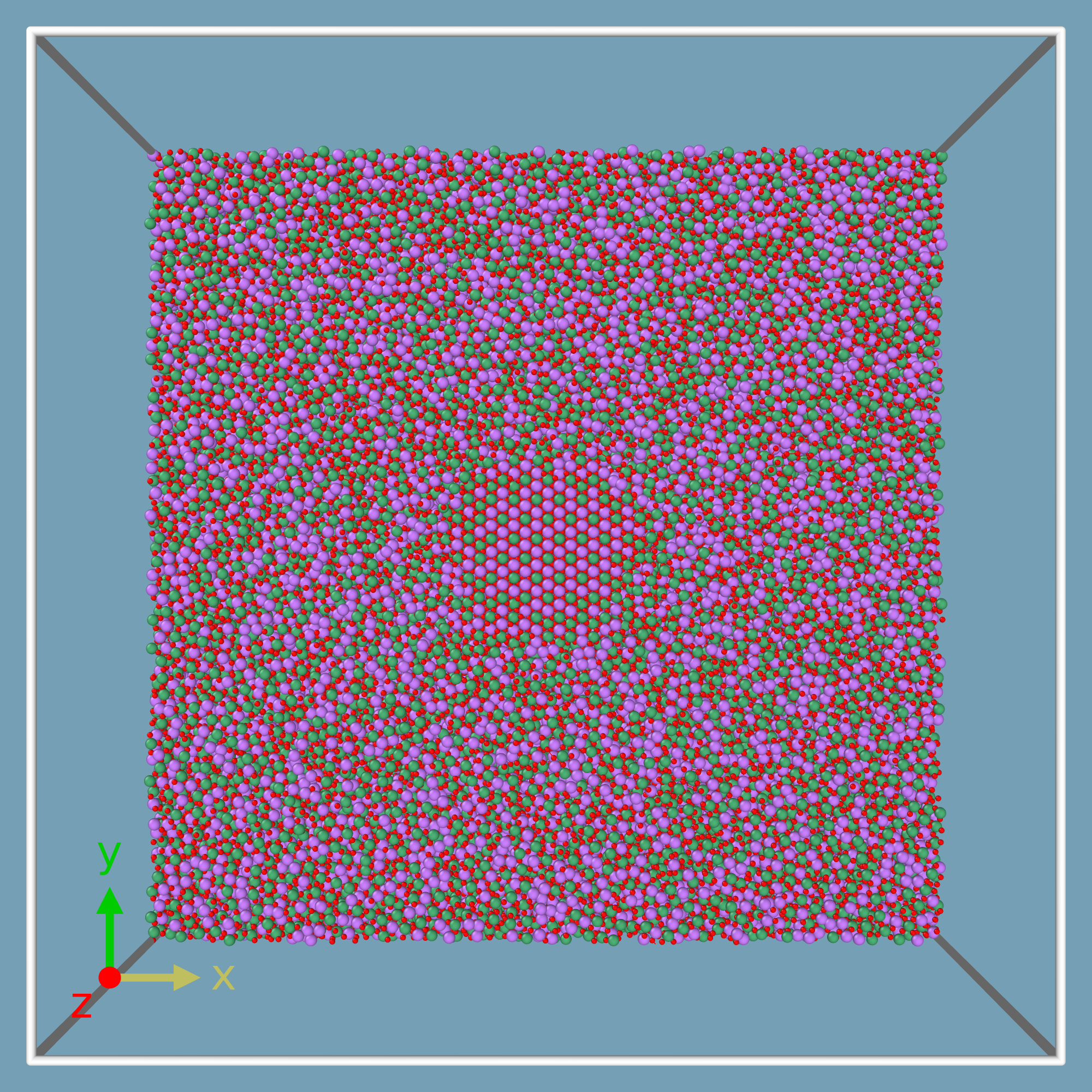}
            \end{minipage}
		\begin{minipage}[b]{.47\textwidth}
			\centering
			\includegraphics[width=0.93\textwidth]{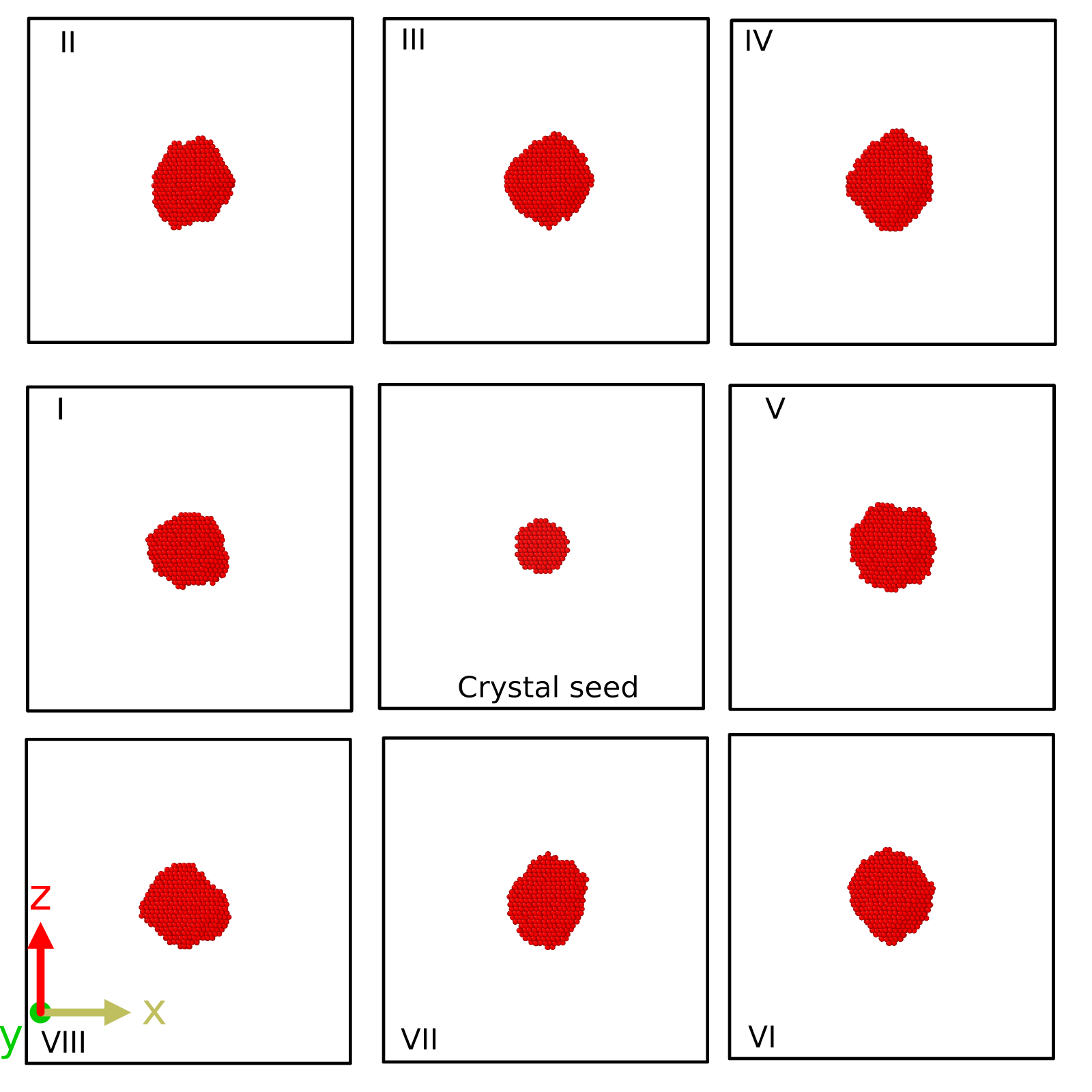}
            \end{minipage}
            \caption{Structure of a representative seeded glass model (left) and ML algorithm predicted crystal-like atoms at the end of the MQ cycle.}
            \label{fig:struct_models}
\end{figure*}
\section{\label{sec:level3}Results and Discussion}
\subsection{LNO}
\subsubsection{\label{sec:level3A}Structure of seeded glass}
\begin{figure*}[!ht]
            \begin{minipage}[b]{.42\textwidth}
			\centering
			\includegraphics[width=0.99\textwidth]{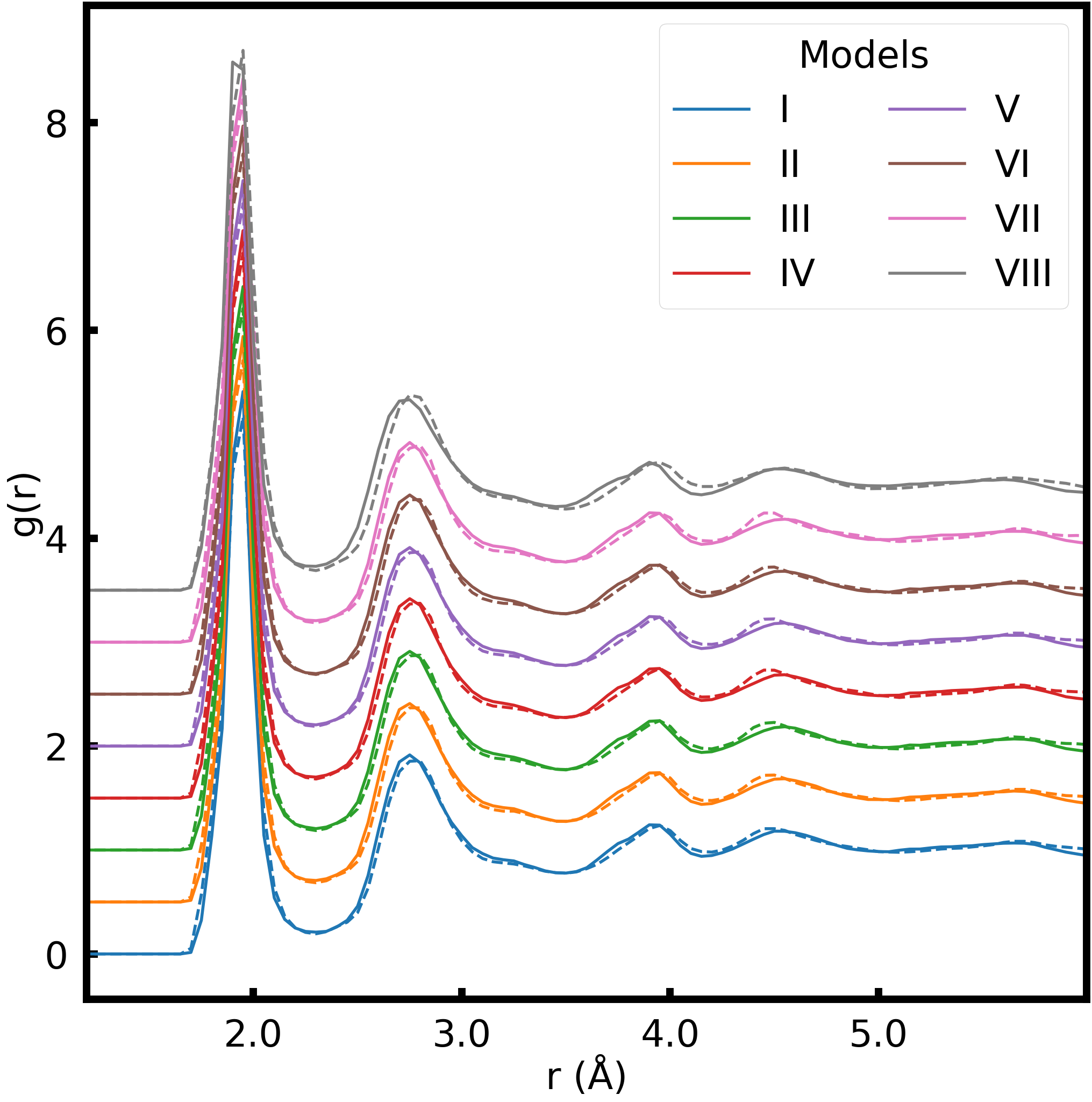}
            \end{minipage}
            \begin{minipage}[b]{.53\textwidth}
			\centering
			\includegraphics[width=0.99\textwidth]{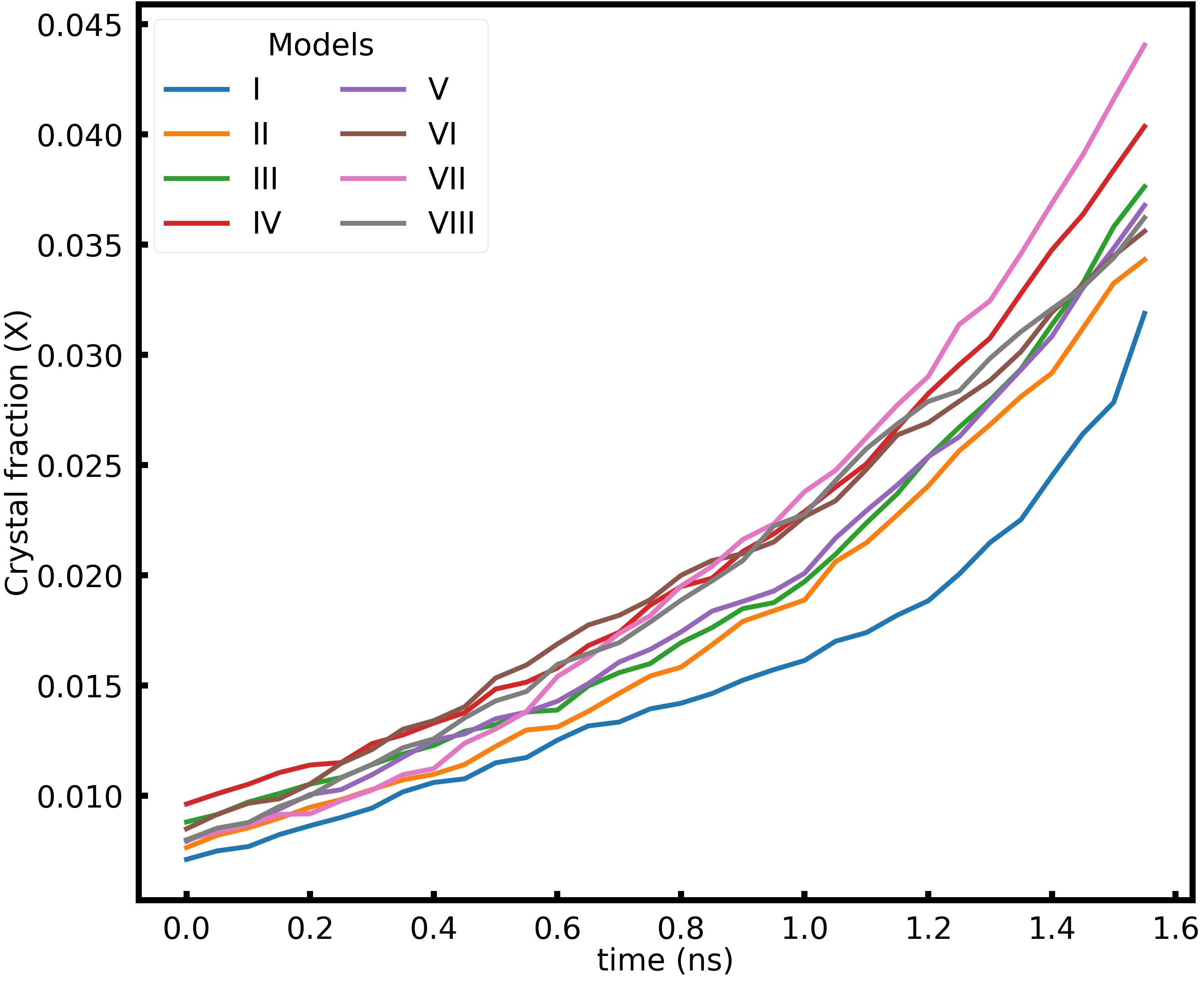}
            \end{minipage}
            \caption{Radial distribution function of seeded glass models after the melt-quench cycle (bold lines) and after SCG (dashed lines)(left panel). Evolution of the crystal fraction (X) during SCG in different models (right panel).}
                \label{fig:growth_rate}
\end{figure*}

The final structure obtained from the melt-quench cycle has a crystal seed surrounded by a glassy arrangement of atoms. The structure of a representative model of seeded glass is shown in Fig.~\ref{fig:struct_models} (left). The seeded glass ensembles underwent different melt-quench cycles thereby creating a different environment around the seed as shown in Fig.~\ref{fig:struct_models} (right). The radial distribution function (RDF) for seeded glass models, shown as bold lines in Fig.~\ref{fig:growth_rate} (left), revealed an insignificant qualitative difference. This suggests similar metal-oxygen (first peak), oxygen-oxygen (second peak), and other possible atomic environments across all the models. The overall structural information of the nearest neighbors and the second nearest neighbors captured by the RDF is essentially independent of the simulation time, provided the simulation is long enough. Such findings confirm the reliability of the melt-quench algorithm employed and the uniformity of the glass outside the seed in terms of structural correlations.


\begin{figure*}[!ht]
            \begin{minipage}[b]{.48\textwidth}
			\centering
			\includegraphics[width=0.93\textwidth]{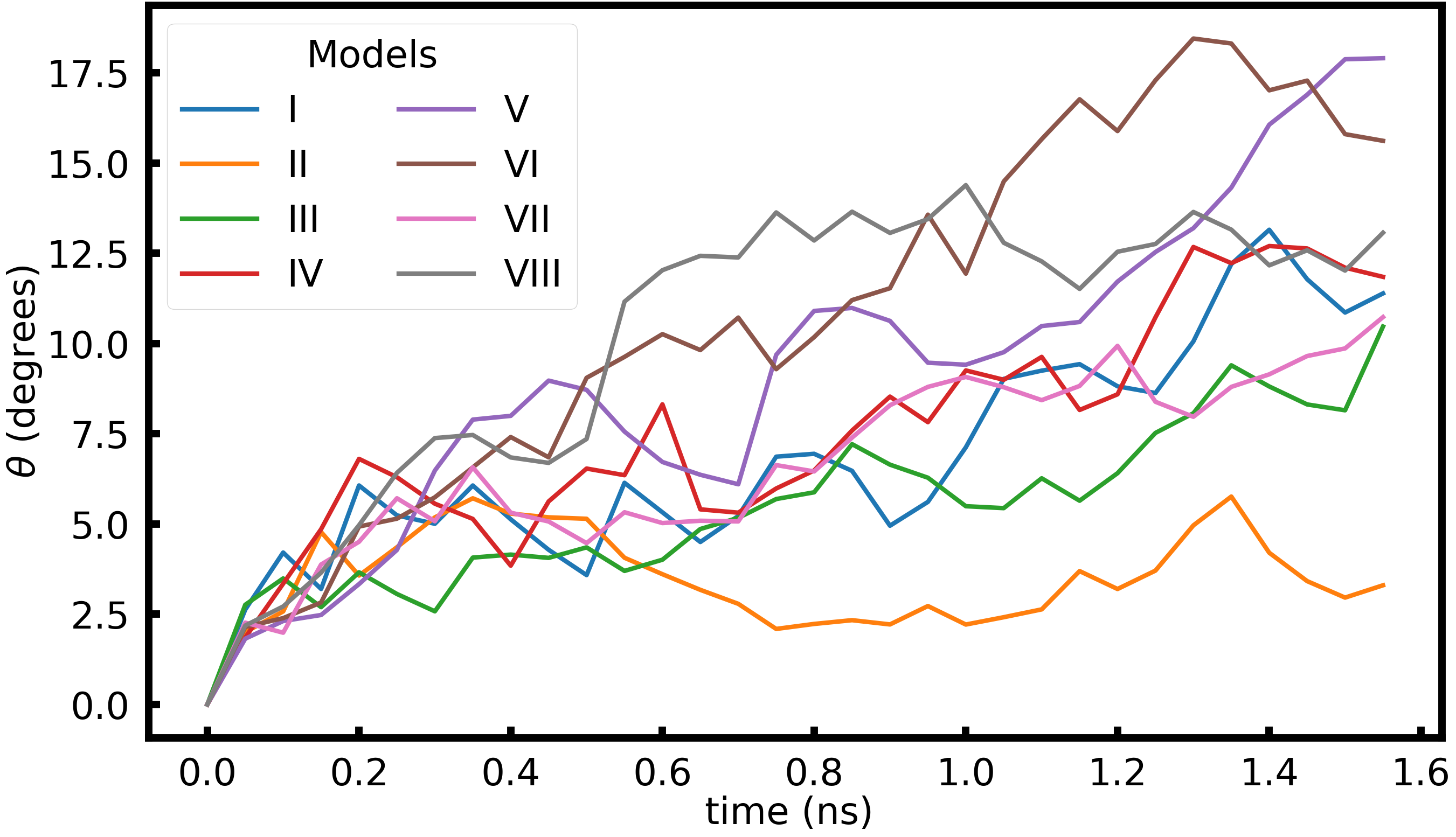}
            \end{minipage}
		\begin{minipage}[b]{.47\textwidth}
			\centering
			\includegraphics[width=0.93\textwidth]{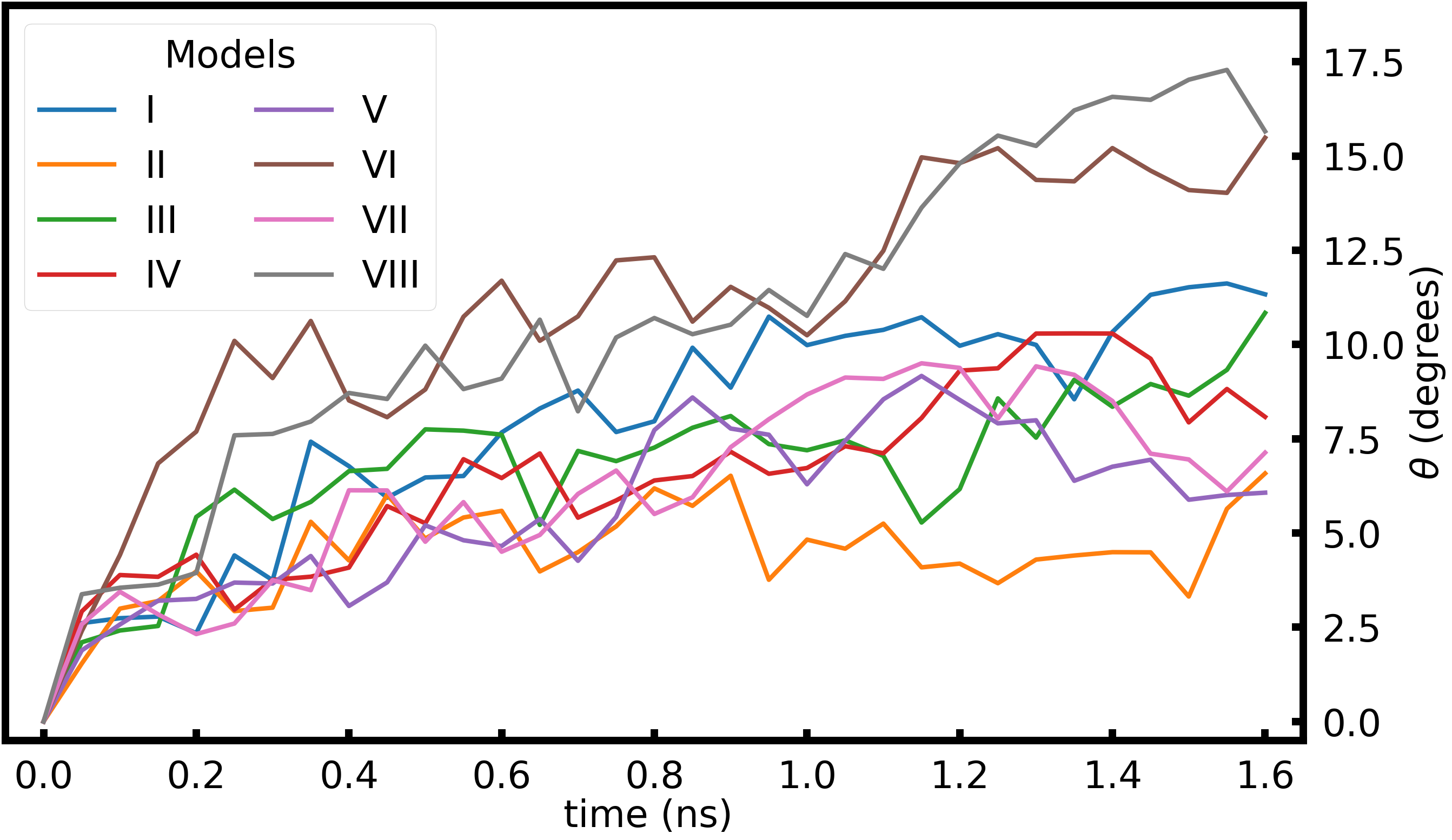}
            \end{minipage}
            \caption{Comparison of the time evolution of crystal rotation across various LNO models for NPT (left) and NVT (right) ensemble during SCG.}
            \label{fig:rot_NPT_NVT}
\end{figure*}
\begin{figure*}[!ht]
            \begin{minipage}[b]{.48\textwidth}
			\centering
			\includegraphics[width=0.93\textwidth]{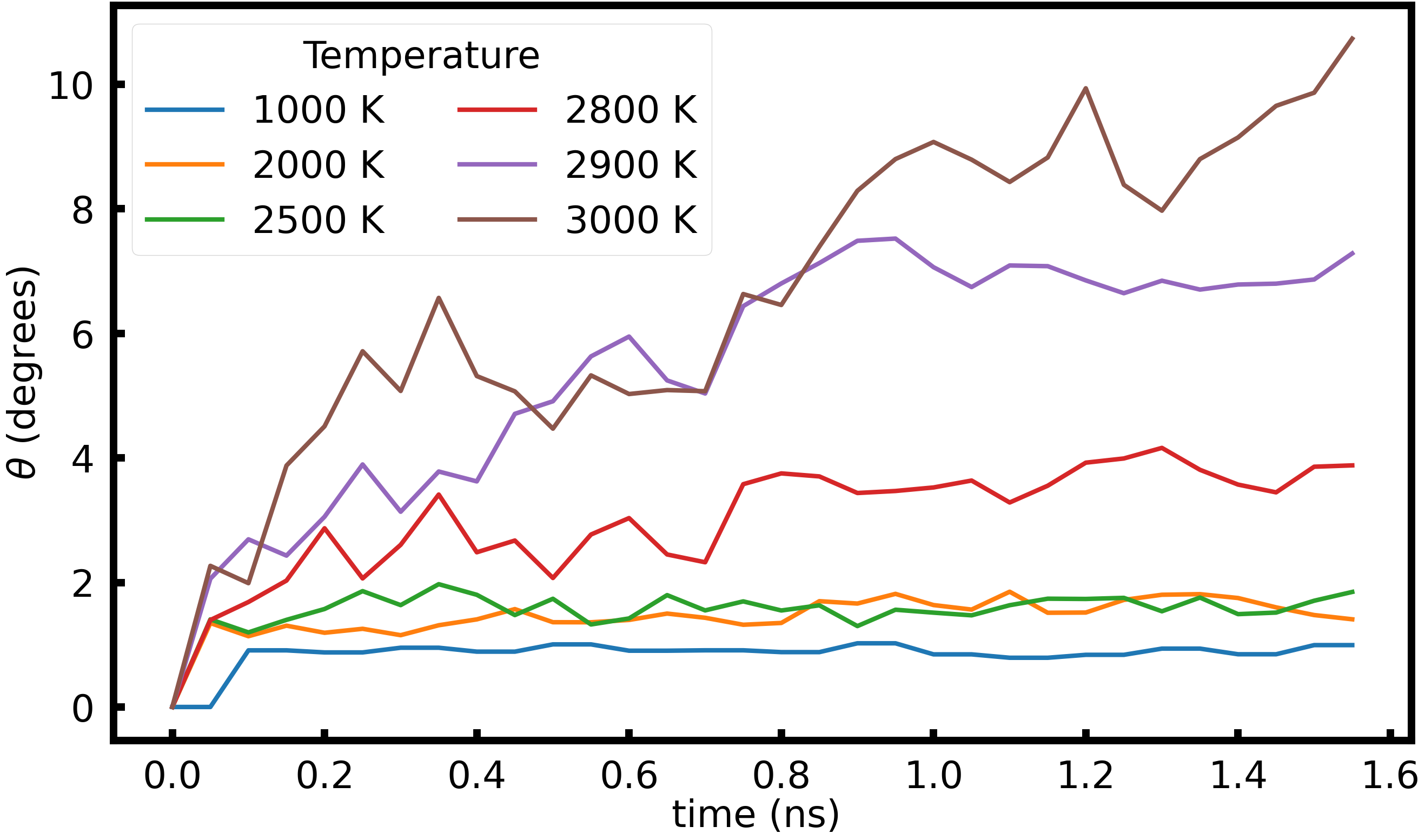}
            \end{minipage}
            \caption{Comparison of the evolution of crystal rotation under the NPT ensemble at various temperatures for Model VII.}
            \label{fig:rot_vs_T}
\end{figure*}

\subsubsection{\label{sec:level3B}Structural change due to SCG}
The RDF of the system after SCG, shown as dashed lines in Fig.~\ref{fig:struct_models} (left), suggests an improved ordering over the short and intermediate length scale compared to the seeded glass. This finding spans across all the models created in this study, corroborating the authenticity of the simulation algorithm used. The slight shift in each RDF peak to higher distances can be attributed to the overall increase in volume during the NPT ensemble implemented during the SCG routine.


\subsubsection{\label{sec:level3C}Growth rate across varying melt quench routine}

The eight different models of seeded LNO glass we prepared had a different atomic arrangement immediately outside the seed, although the average structure in the short- and medium-range order is very similar. The quantification of crystallinity in these models during SCG is tracked with the evolution of the fraction of crystal-like atoms. The identification of whether an atom is crystal or glass is made using a machine-learned clustering algorithm based on static structural features (Steinhardt parameter~\cite{Steinhardt}, number of connections~\cite{tenWolde}) and temperature dynamics of the atoms~\cite{RAJ2023}. The time evolution of the crystal fraction over various models, shown in Fig.~\ref{fig:growth_rate}, suggests that the growth rate follows almost similar paths except for Model I, which may be attributed to the short melt equilibration time employed during the melt-quench step.


\subsubsection{\label{sec:level3D}Crystal rotation during SCG}
We have used the technique discussed in Sect.~\ref{sec:level2B} to study the time evolution of the rotation of the growing crystal relative to its original orientation. This analysis has been repeated for all the models (I-VIII) that have a different glass environment around the seed. The time evolution of the rotation for various models, shown in Fig.~\ref{fig:rot_NPT_NVT}, indicates that seed rotation is observed across all models subjected to SCG. This finding emphasizes two key points: first, the glass structure surrounding the seed plays a crucial role in determining the nature of the crystal’s rotation. Second, the glass applies nonuniform forces on the seed’s surface, inducing its rotation, which implies that the glass cannot be considered isotropic at the length scale of the glass-seed  interface. This seed rotation ceases after some time when the curve flattens out and the crystal has grown large enough to resist the torque on it due to the surrounding glass. The exact point where the curve becomes flat may vary from one glass structure to another. The varying nature of rotation among different models suggests that the rotation is a function of the glass structure surrounding it and not the seed itself.

To gain a better understanding of the observed seed rotation, we studied the structural changes during seeded crystal growth in more depth.
\subsubsection{\label{sec:level3C2}Crystal Rotation vs volume/density change}
Our SCG simulations were performed using an NPT ensemble where the box boundary was free to change in order to maintain constant pressure conditions. Consequently, in all our NPT simulations, the volume of the system increased, which corresponded to a decrease in density. To explore the effect of increasing volume (decreasing density) on the observed rotation, we also performed an SCG simulation under the NVT ensemble and compared the resulting rotation behavior with that from the NPT ensemble. The time evolution of rotation in the NVT ensemble, shown in Fig.~\ref{fig:rot_NPT_NVT} (right), indicates that rotational motion is present throughout the glass structures surrounding the seed. Notably, the overall rotational behavior observed in both NVT and NPT ensembles is similar, strongly suggesting that the rotation originates from the surrounding glass structure rather than being an artifact of the annealing method used in SCG.

\subsubsection{\label{sec:level3C1}Crystal Rotation vs temperature}
SCG at 3000 K certainly induces large atomic displacements in the system because it is close to the melting temperature. At such high temperatures, atoms gain high velocities, the magnitude of which varies inversely to the atomic mass. This could lead to nonuniform forces acting along the surface of the seed, thereby inducing a net torque that causes it to rotate. To study the extent of the effect of temperature on the observed crystal rotation, we examined the time evolution of crystal rotation for various temperatures as shown in Fig.~\ref{fig:rot_vs_T}. It suggests that the magnitude of the rotation is directly proportional to the temperature. It could be attributed to the fact that the magnitude of non-uniform forces exerted by the glass on the crystal seed is proportional to the temperature.



\subsubsection{\label{sec:level3C3}Crystal Rotation vs glass density}
First of all, density fluctuations must be understood. Because the crystal seed has a density higher than that of the surrounding glass, the seed atoms that were frozen during the MQ cycle tend to free themselves up by expanding. This happens because the seed tries to match itself with the density of the glass around it. This causes a slight change in the bond lengths, but the crystal seed does not lose its periodic arrangement of atoms. The extent of the effect of the inherent proclivity of the seed to expand during the early stage of the SCG on the nature of rotation can be understood by modeling the effect of a change in the density of the glass surrounding the crystal on the observed rotation. For this, we created seeded glass where the density of the glass outside the seed is $4.3 \ \textrm{g/cc}$, Model VIII, much closer to the density of the crystal seed $4.4\ \textrm{g/cc}$. Our results for the crystal growth rate, shown in Fig.~\ref{fig:growth_rate} (right), and the evolution of the rotation during SCG, shown in Fig.~\ref{fig:rot_NPT_NVT}, suggest that the crystal rotation and growth seen in LNO are not a consequence of the density difference between glass and crystal.


\begin{figure*}[!ht]
            \begin{minipage}[b]{.48\textwidth}
			\centering
			\includegraphics[width=0.93\textwidth]{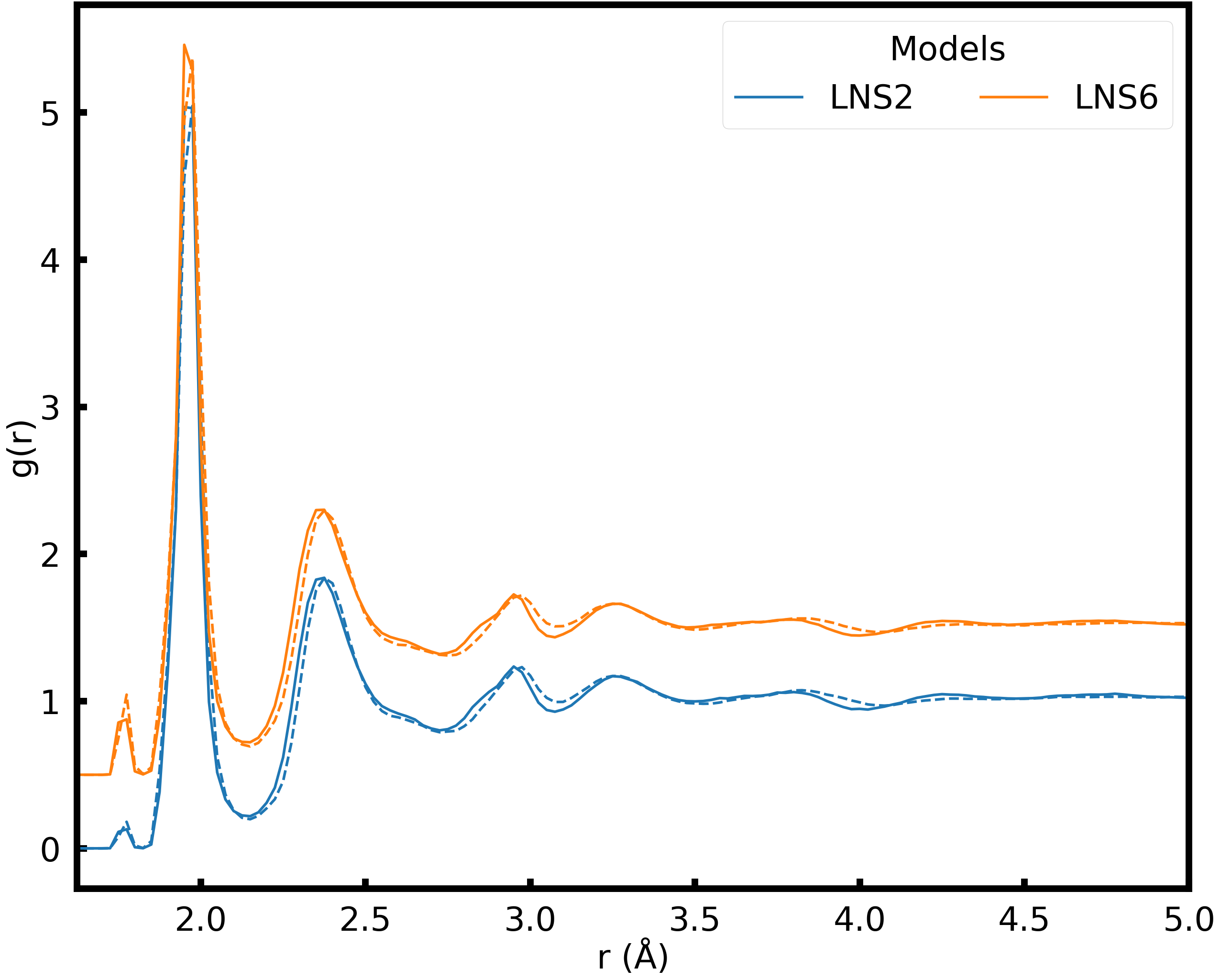}
            \end{minipage}
            \caption{Radial distribution function (right) of seeded glass models after the melt-quench cycle (bold lines) and after SCG (dashed lines). }
            \label{fig:RDF_LNS}
\end{figure*}

\begin{figure*}[!ht]
            \begin{minipage}[b]{.48\textwidth}
			\centering
			\includegraphics[width=0.93\textwidth]{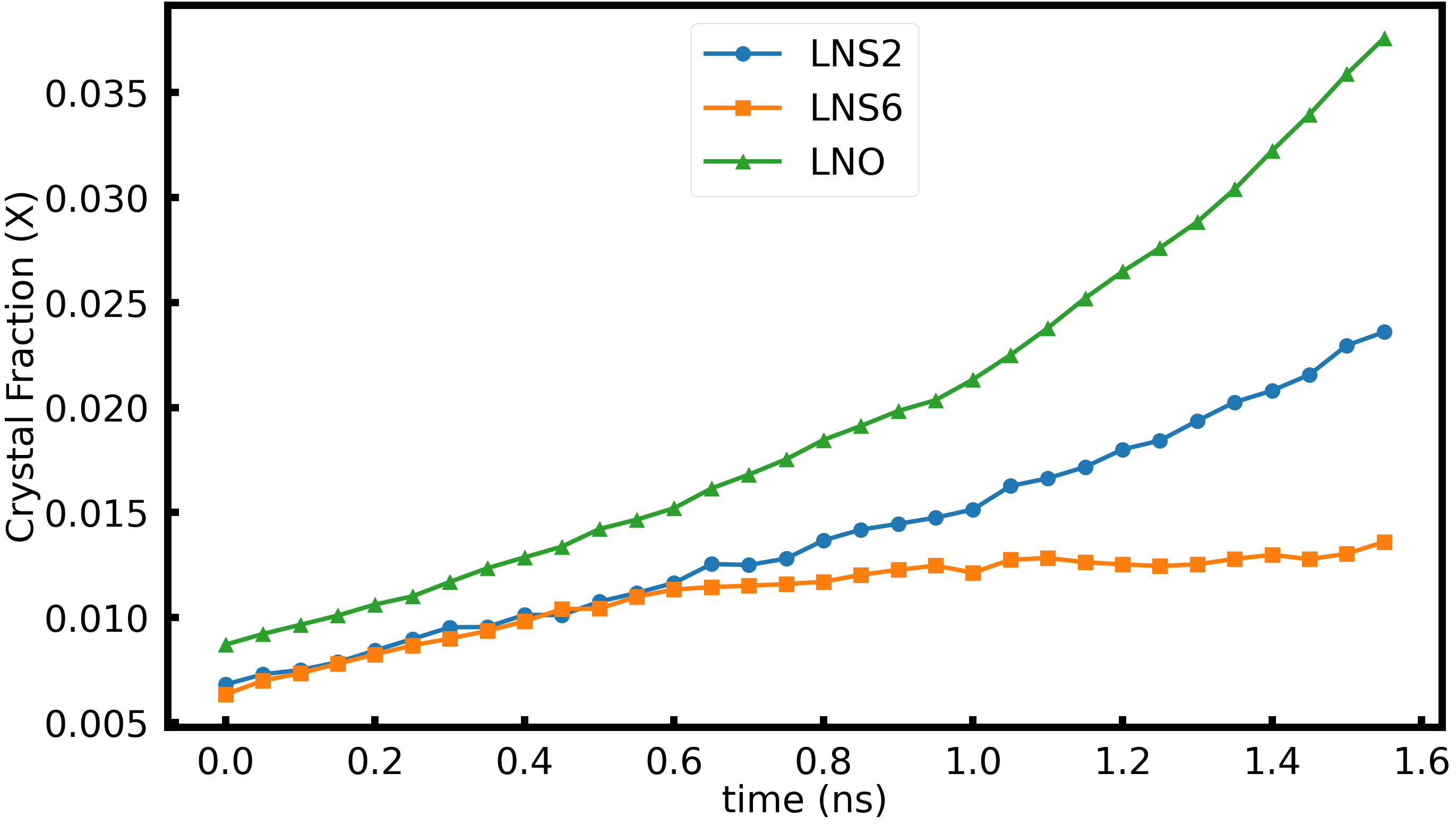}
            \end{minipage}
		\begin{minipage}[b]{.47\textwidth}
			\centering
			\includegraphics[width=0.93\textwidth]{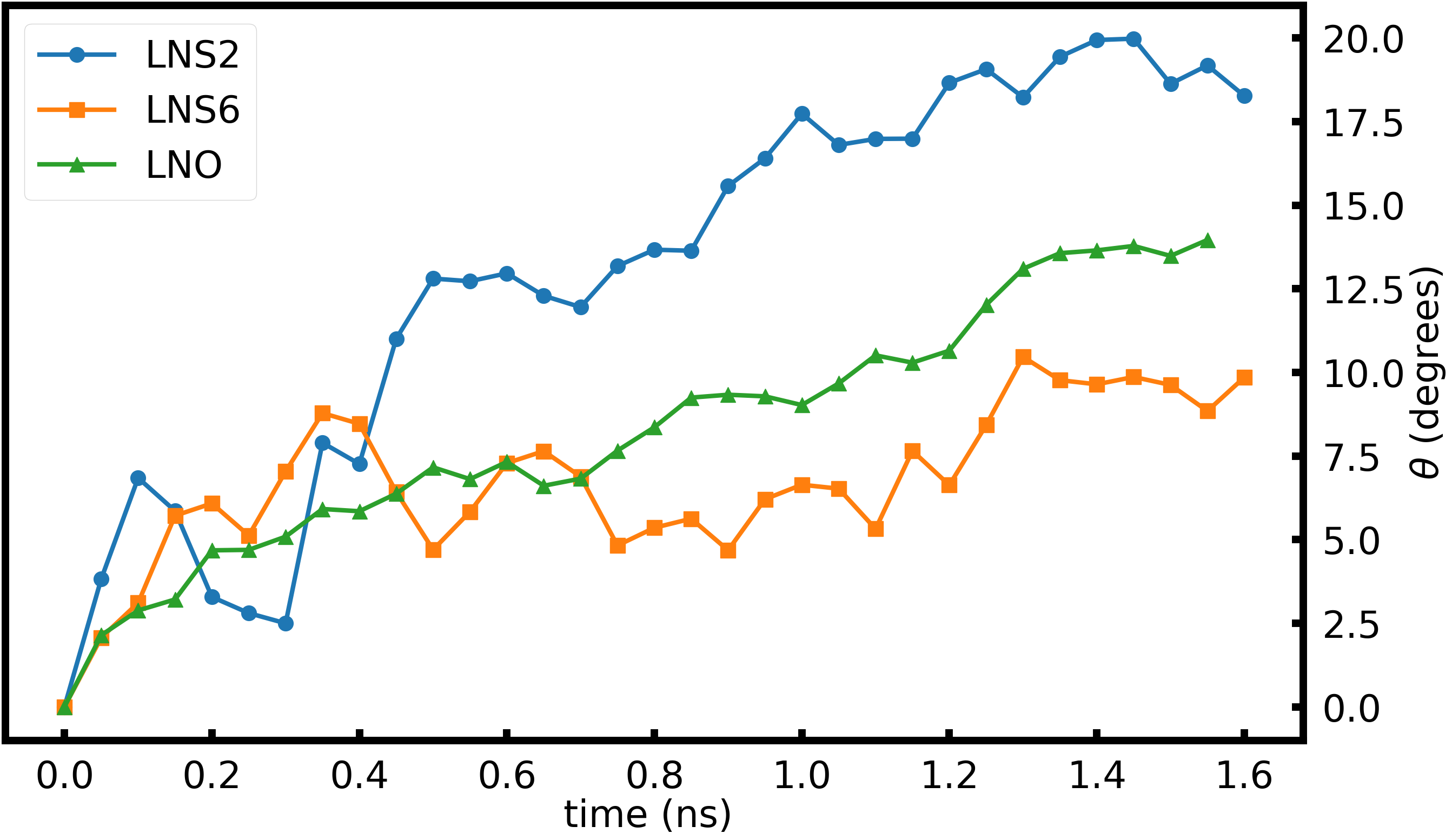}
            \end{minipage}
            \caption{Evolution of crystal fraction (left) and rotation magnitude (right) during SCG of LNO crystal in LNS glass compared with SCG in LNO glass with similar cooling cycle.}
            \label{fig:LNS2_vs_LNS6}
\end{figure*}
\begin{figure*}[!ht]
            \begin{minipage}[b]{.95\textwidth}
			\centering
			\includegraphics[width=0.93\textwidth]{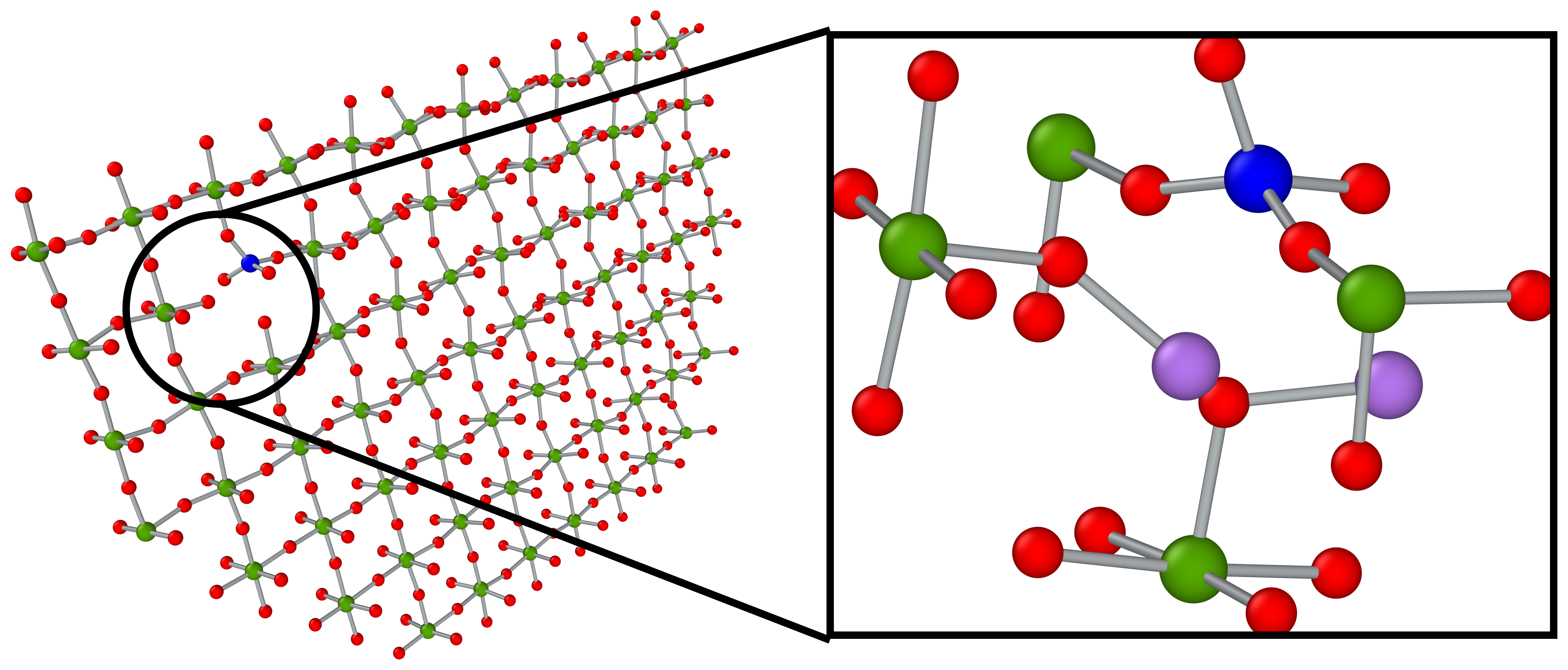}
            \end{minipage}
	
            \caption{A Nb (green) atom plane with O atom (red) containing Si atom (blue) inside the grown crystal (left). Zoomed in image of the region with a defect shown with all atomic species present (right). Li atoms are colored purple}
            \label{fig:Si_in_xtal}
\end{figure*}

\begin{figure*}[!ht]
            \begin{minipage}[b]{.60\textwidth}
			\centering
			\includegraphics[width=0.93\textwidth]{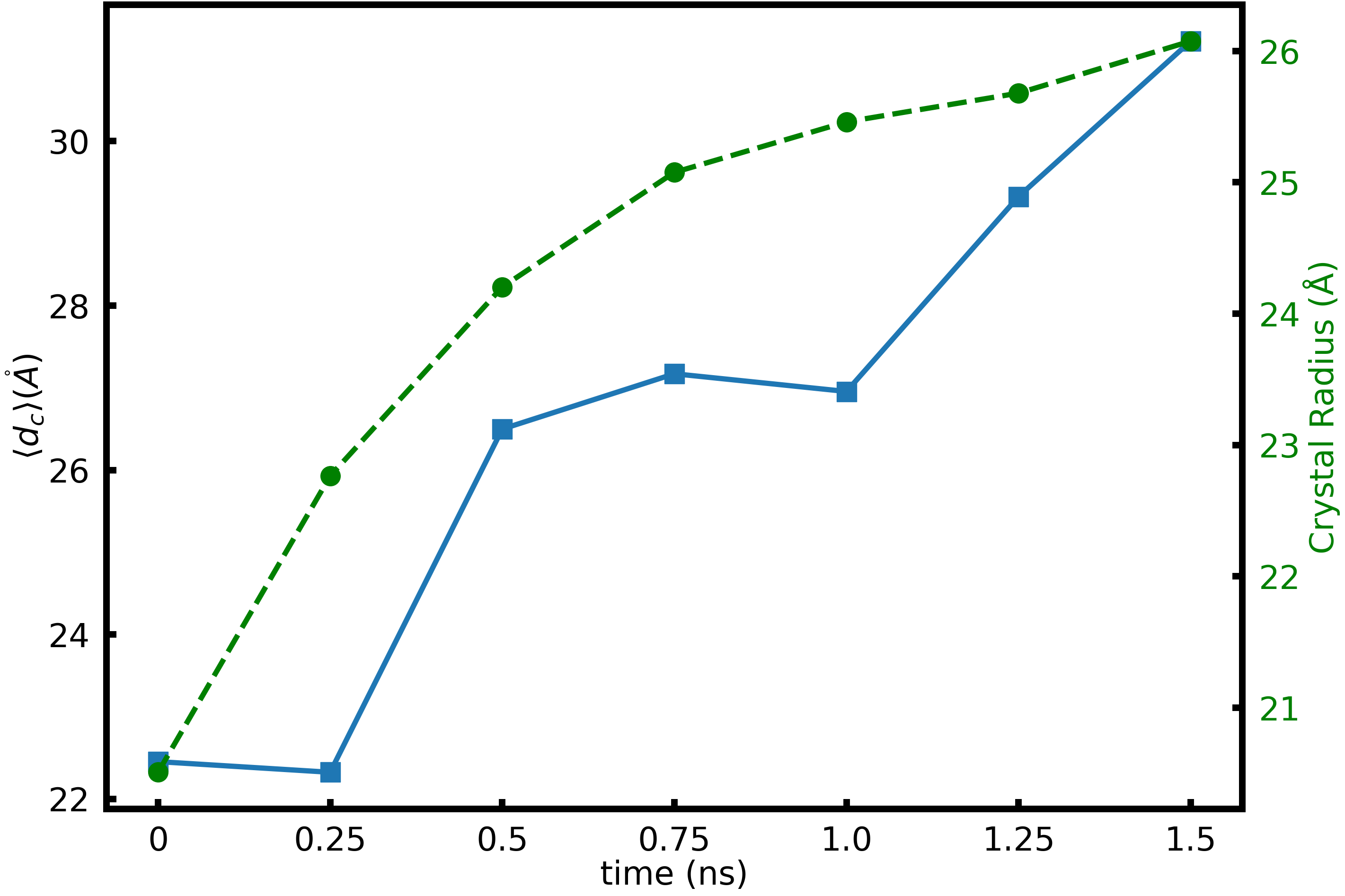}
            \end{minipage}
	
            \caption{Average distance of 100 closest Si atoms to the center of the seed in LNS6 (solid line) glass and the calculated crystal size (dashed line) as a function of SCG time}
            \label{fig:Si_dist}
\end{figure*}

\subsection{LNS2 and LNS6}
\subsubsection{\label{sec:level4A} Structure of the seeded glass}
The structure of the lithium niobosilicate glasses obtained after the melt-quench cycle showed an amorphous nature characterized by the RDF as seen in Fig.~\ref{fig:RDF_LNS}. The peak at the lowest separation, at $1.65\ \mathring{A}$, corresponds to Si-O bonds~\cite{Sun2022}. The nearest neighbor environment around individual cations shows that Si has a tetrahedral coordination forming SiO\textsubscript{4} units as reported in various silica containing glasses both experimentally~\cite{Galeener1983} and theoretically~\cite{BHATTARAI2016}. This result suggests that the Si-O interaction used in the simulations produces expected results and validates the parameters used for the study of SCG in LNS glasses.


\subsubsection{\label{sec:level4B} Crystal Growth Rate and Rotation}
After being subjected to SCG conditions, NPT ensemble at 3000 K, the crystal seed continues to grow, advancing the interface into the surrounding glass. The seed growth rate is tracked using the same machine learning-based algorithm as was used for LNO and is plotted in Fig.~\ref{fig:LNS2_vs_LNS6}. The figure shows that the crystal growth rate decreases significantly as SiO\textsubscript{2} is gradually added to LNO composition, to make LNS2 and LNS6 glasses. This may be attributed to various factors. First, the nearest-neighbor environment around Li or Nb atoms differs from that of Si. Secondly, silica, a glass former, is known to reduce the atomic movements around it because of the inherent higher viscosity.


The rotation of the seeds during SCG is presented in Fig.~\ref{fig:LNS2_vs_LNS6} (right). We observe that the seed rotated during SCG in both LNS2 and LNS6. Interestingly enough, we have seen lower magnitude of rotation in LNS6 compared to LNS2 which may be attributed to the higher silica concentration slowing down the atomic dynamics and consequently the crystal growth rate. However, the crystal rotation is not significantly changed by the presence of Si, suggesting that the factors governing rotation and those driving growth are not the same. The former is determined by the heterogeneity of glass structure around the seed, whereas the latter is related to the mobility of slowest moving atoms across the growth interface. The crystal growth process has a rotational aspect coupled with the addition of atoms to the growing crystal seed. This is in line with findings from pure LNO glass above.


\subsubsection{\label{sec:level4C} Si dynamics during SCG}
One important aspect of the LNO crystal growing inside the LNS glass is the effect of Si atoms on the crystallization process. Our finding suggests that increasing the silica content significantly reduces the growth rate of the LiNbO\textsubscript{3} crystal in the LNS glass. XRF studies on laser crystallized LNO crystals in LNS glass have shown Si incorporation into the growing crystal~\cite{barker2024femtosecond}. Analysis of the composition of the growing seed shows Si incorporation, substituting into the Nb plane, as shown in Fig.~\ref{fig:Si_in_xtal} (left). However, there appear to be no Si atoms in the part of the seed that was originally inserted. This could be attributed to the fact that the creation of a Si defect in the crystal seed with otherwise perfect lattice is not energetically favored. The Si atoms in the LNO crystal have a low solubility and their diffusivity is much slower than the kinetics of crystallization. The rigid skeleton of the crystal is mostly tied to the heavier Nb atoms. Most of the Si atoms that we see in the crystal grown from the seed during SCG are the ones that were present in the path of the growing crystal and were engulfed by the growing crystal. The Si atoms inside these grown crystals have a tetrahedral arrangement of O atoms around them suggesting the presence of a disorder in the atomic arrangement around the Si atoms and thus the nearby Nb or Li atoms. A zoomed-in view of the defected region shown in Fig.~\ref{fig:Si_in_xtal} (right) suggests that there are no dangling bonds even in the defected region. The nature of the Si ion movements and their distribution within the system can be ascertained from the time evolution of the distance of the 100 closest Si atoms from the seed during SCG. It is shown in Fig.~\ref{fig:Si_dist}. This analysis suggests that overall Si atoms near the seed appear to be being pushed out by the growing crystal as its size expands. Thus the LNO crystal grown inside the LNS glass consists of some Si defects on Nb sites with most of the Si atoms pushed into the residual glass. A visual interpretation of the movement of these 100 closest Si atoms during the SCG has been shown in a movie~\href{https://drive.google.com/file/d/1NZ8jyqTN44777GeFdJBRQpF5NcYuZ5WG/view?usp=drive_link}{here}.

\section{\label{sec:level5}Conclusions}
We have simulated the SCG of LNO in both pure LNO and LNS glasses using systems comprising approximately half a million atoms. Our simulations discovered that in the early stages of growth, the seed embedded in the glass rotates as a whole until it becomes too large to rotate. By varying the structure of the glass surrounding the seed, we find that the surrounding glass significantly influences the nature of the crystal rotation. The glass imposes non-uniform forces on the seed, inducing rotation and revealing that it cannot be regarded as isotropic or uniform at the length scale of the glass–crystal interface. Importantly, the observed seed rotation is not a simulation artifact, as it persists across different simulation conditions (NPT and NVT ensembles), temperatures, and glass densities.  

We also investigated the impact of silica concentration in the LNS glass on both the growth rate and rotation dynamics of the LNO seed during SCG. As expected from experiments, a higher fraction of the network former SiO\textsubscript{2} suppresses seed crystallization. Moreover, the LNO crystals grown in LNS glass push out Si atoms but also incorporate some of them, introducing local disorder due to the mismatch between the nearest-neighbor environments of Si and those of Nb or Li atoms. 

\section{\label{sec:level5}Acknowledgments}
This work was supported by the US Department of Energy (DoE),
Office of Science, Basic Energy Sciences, under award DE-SC0005010. Computer simulations were performed using ACCESS~\cite{ACCESS} at Pittsburgh Supercomputing
Center through mat230020p allocation from the Advanced Cyberinfrastructure
Coordination Ecosystem: Services \& Support (ACCESS) program, which is
supported by the National Science Foundation, United States grants 2138259, 2138286, 2138307, 2137603, and 2138296. We also acknowledge the computing resources available
through Lehigh Supercomputer center to perform some of our simulations. The authors thank Dr. Jincheng Du at UNT for guidance with molecular dynamics simulation setup in LAMMPS.

\newpage
\bibliography{reference}
\end{document}